\documentclass[11pt]{article}
\usepackage{amssymb}
\usepackage{epsfig}

\newcommand{\BE}{\begin{equation}}
\newcommand{\EE}{\end{equation}}
\begin{document}
\begin{titlepage}

\vspace*{1mm}
\begin{center}

   {\LARGE{\bf Motion toward the Great Attractor \\
from an ether-drift experiment }}

\vspace*{14mm}
{\Large  M. Consoli and E. Costanzo}
\vspace*{2mm}\\
{\small
Istituto Nazionale di Fisica Nucleare, Sezione di Catania \\
Dipartimento di Fisica e Astronomia dell' Universit\`a di Catania \\
Via Santa Sofia 64, 95123 Catania, Italy \\} \vspace*{6mm}
{\Large V. Palmisano } \vspace*{2mm}\\
{\small Dipartimento di Fisica Nucleare dell'Universit\`a di
Messina\\
Istituto Nazionale di Fisica Nucleare, Gruppo Collegato di Messina\\
Salita Sperone 31, 98166 Messina, Italy}
\end{center}
\begin{center}
{\bf Abstract}
\end{center}
Since the end of 80's, the region of sky of galactic coordinates
($l\sim 309^o$, $b\sim 18^o$), corresponding to a declination
$\gamma\sim -44^o$ and right ascension $\alpha\sim 202^o$, usually
denoted as the "Great Attractor", is known to control the overall
galaxy flow in our local Universe. In this sense, this direction
might represent a natural candidate to characterize a hypothetical
Earth's "absolute motion". Our analysis of the extensive ether-drift
observations recently reported by an experimental group in Berlin
provides values of $\alpha$ and $\gamma$ that coincide almost
exactly with those of the Great Attractor and {\it not} with the
values $\gamma\sim -6^o$ and $\alpha\sim 168^o$ obtained from a
dipole fit to the anisotropy of the CMB. This supports in a new
fashion the existence of a discrepancy between the observed motion
of the Local Group and the direction obtained from the CMB dipole.

Submitted to the Astrophysical Journal
\end{titlepage}

\section{Introduction}

Since the end of 80's, a region of the sky in the direction of the
Centaurus cluster, the so called "Great Attractor", at galactic
coordinates ($l\sim 309^o$, $b\sim 18^o$), corresponding to a
declination $\gamma\sim -44^o$ and right ascension $\alpha\sim
202^o$, is known to play an important role in describing the
deviations of galaxies from a pure Hubble flow in our local
Universe,
\cite{lynden88,dressler88,faber,burstein,erratum,attractor}. In this
sense, the Great Attractor, marking a preferred direction in space,
might represent a natural candidate to characterize a hypothetical
Earth's "absolute motion".

On the other hand, since the discovery of an anisotropy in the
cosmic microwave background (CMB), it has been generally accepted
that the kinematical parameters for such an absolute motion should
coincide with those deduced from a dipole fit to the COBE data
\cite{cobe}. In this context, one predicts a velocity of the Solar
System $v\sim 370$ km/s, an average declination angle $\gamma\sim
-6^o$ and a right ascension $\alpha\sim 168^o$. These values have
usually been adopted in the interpretation of the data from the
ether-drift experiments in a laboratory.

However, the latest ether-drift experiments, combining the
possibility of active rotations of the apparatus with the use of
cryogenic optical resonators \cite{schiller,peters,tobar}, have
reached such a high precision (${\cal O}(10^{-16})$ in the relative
frequency shifts) to require a fully model-independent analysis of
the data. In fact, the physical nature of a hypothetical preferred
frame is still unknown. Therefore, assuming from the very beginning
one particular set of  values for $(v,\gamma,\alpha)$ one might
introduce uncontrolled errors in the interpretation of the
experimental results.

This is even more true noticing that a fully model-independent
analysis \cite{noi} of the extensive ether-drift observations
reported by Herrmann et al. in Ref.\cite{peters} provides an average
(absolute) value of the declination angle $|\gamma|\sim 43^o \pm
3^o$ that would rather favour an alternative of the type represented
by the Great Attractor. In this paper we'll further extend the
analysis of Ref.\cite{noi} that, being limited to a restricted set
of observables, could not determine the value of $\alpha$ and the
sign of $\gamma$. As we shall illustrate, our new results, after
inclusion of other observable quantities from Ref.\cite{peters},
provide values of $\alpha$ and $\gamma$ that are in remarkable
agreement with those of the Great Attractor. For this reason, our
results, while providing the first modern evidence for an ether
drift from a laboratory experiment, support the indications obtained
from the observed motion of galaxies.

The plane of the paper is as follows. In Sect.2 we shall report the
relevant formalism used in the analysis of the ether-drift
experiments and the basic experimental data of Ref.\cite{peters}. In
Sect.3 we shall present our analysis of these data while in Sect.4
we shall summarize our results and present our conclusions.
\section{General formalism and experimental data}

The starting point for our analysis is the expression for the
relative frequency shift of the two optical resonators at a given
time $t$. This is expressed as \cite{peters}
\BE
\label{basic2}
      {{\delta \nu (t)}\over{\nu_0}}   =    {S}(t)\sin 2\omega_{\rm rot}t +
      {C}(t)\cos 2\omega_{\rm rot}t
\EE
where $\omega_{\rm rot}$ is the rotation frequency of one resonator
with respect to the other which is kept fixed in the laboratory and
oriented north-south. The Fourier expansions of the two amplitudes
$S(t)$ and $C(t)$ are predicted to be
\BE
\label{amorse1}
      {S}(t)   =    {S}_{s1}\sin\tau +{S}_{c1} \cos\tau
       + {S}_{s2}\sin(2\tau) +{S}_{c2} \cos(2\tau)
\EE
\BE
\label{amorse2}
      {C}(t) = {C}_0 +
      {C}_{s1}\sin\tau +{C}_{c1} \cos\tau
       + {C}_{s2}\sin(2\tau) +{C}_{c2} \cos(2\tau)
\EE
where $\tau=\omega_{\rm sid}t$ is the sidereal time of the observation
in degrees and $\omega_{\rm sid}\sim {{2\pi}\over{23^{h}56'}}$.

Introducing the colatitude of the laboratory $\chi (\sim 37.5^o$ for
Berlin), one finds the expressions reported in Table I of
Ref.~\cite{peters},
\BE
\label{C0}
      {C}_0 =-K
      {{\sin^2\chi}\over{8}} (3 \cos 2{\gamma} -1) ,
\EE
\BE
\label{CS1}
      {C}_{s1}= {{1}\over{4}}K
      \sin 2{\gamma} \sin{\alpha} \sin 2\chi ,
\EE
\BE
\label{CC1}
      {C}_{c1}={{1}\over{4}} K \sin 2{\gamma}
      \cos{\alpha} \sin 2\chi ,
\EE
\BE
\label{CS2}
      {C}_{s2} = {{1}\over{4}}K \cos^2{\gamma}
      \sin2{\alpha}  (1+ \cos^2\chi) ,
\EE
\BE \label{CC2}
      {C}_{c2} = {{1}\over{4}}K \cos^2{\gamma}
      \cos2{\alpha} (1+ \cos^2\chi)
\EE
where \BE \label{cappa}
 K=(1/2-\beta+\delta) {{ v^2 }\over{c^2}} \EE
and $(1/2-\beta+\delta)$ indicates the Robertson-Mansouri-Sexl
\cite{robertson,mansouri} (RMS) anisotropy parameter. The
corresponding $S-$quantities are also given by
${S}_{s1}=-{C}_{c1}/\cos\chi$, ${S}_{c1}={C}_{s1}/\cos\chi$,
${S}_{s2}= -{{2\cos\chi}\over{1+\cos^2\chi}}{C}_{c2}$ and ${S}_{c2} = {{2\cos\chi}\over{1+\cos^2\chi}}{C}_{s2}$. It might be interesting
that these relations can also be derived from an old paper by Nassau
and Morse, published in the Astrophysical Journal about eighty years
ago (see Eqs.(20-24) of Ref.\cite{nassau}).

Notice a critical detail for the separation of the signal in its
elementary components. A small mismatch in the definition of the
sidereal time, say $\tau^{\rm true}=\tau +\Delta \tau$, induces a
rotation of the various parameter pairs of angles $\Delta\tau$ and
$2\Delta\tau$ with a corresponding re-definition of the right
ascension $\alpha_{\rm true}= \alpha +\Delta\tau$.

 The experimental data reported in Ref.\cite{peters} were
obtained during 15 short-period observations performed from December
2004 to April 2005. As suggested by the same authors, it is safer to
concentrate on the observed time modulation of the signal, i.e. on
the quantities ${C}_{s1},{C}_{c1},{C}_{s2},{C}_{c2}$ and on their
${S}$-counterparts. In fact, the constant components $ \bar{C}={C}_0
$ and $\bar { S}\equiv S_0$ are most likely affected by spurious
systematic effects such as thermal drift (see also the discussion in
Ref.~\cite{schiller}).

The individual determinations of the various parameters, for each of
the 15 short-period observations, as extracted from Fig.3 of
Ref.\cite{peters}, are reported in Table 1 and Table 2.

\begin{table*}
\caption{The experimental values of the $C-$coefficients for the 15
observation periods of Ref.\cite{peters}. }
\begin{center}
\begin{tabular}{cllll}
\hline\hline Observation~$i$&$C_{s1}[{\rm x}10^{-16}]$ &
$C_{c1}[{\rm x}10^{-16}]$ & $C_{s2}[{\rm x}10^{-16}]$ &
$C_{c2}[{\rm x}10^{-16}]$   \\
\hline
1&$-2.7\pm4.5$ &  $5.3\pm 4.8$ &  $-3.2\pm4.7$ &  $1.2\pm 4.2$ \\
2&$-18.6\pm6.5$ &  $8.9\pm 6.4$ &  $-11.4\pm6.5$ &  $-5.0\pm 6.4$ \\
3&$-0.7\pm3.9$ &  $5.3\pm 3.6$ &  $5.0\pm3.5$ &  $1.6\pm 3.8$ \\
4&$6.1\pm4.6$ &  $0.0\pm 4.8$ &  $-8.1\pm4.8$ &  $-4.0\pm 4.6$ \\
5&$2.0\pm8.6$ &  $1.3\pm 7.7$ &  $16.1\pm8.0$ &  $-3.3\pm 7.2$ \\
6&$3.0\pm5.8$ &  $4.6\pm 5.9$ &  $8.6\pm5.9$ &  $-6.9\pm 5.9$ \\
7&$0.0\pm5.4$ &  $-9.5\pm 5.7$ &  $-5.5\pm5.6$ &  $-3.5\pm 5.4$ \\
8&$-1.1\pm8.1$ &  $11.0\pm 7.9$ &  $0.9\pm8.3$ &  $18.6\pm 7.9$ \\
9&$8.6\pm6.5$ &  $2.7\pm 6.7$ &  $4.3\pm6.5$ &  $-12.4\pm 6.4$ \\
10&$-4.8\pm4.8$ &  $-5.1\pm 4.8$ &  $3.8\pm4.7$ &  $-5.2\pm 4.7$ \\
11&$5.7\pm3.2$ &  $3.0\pm 3.4$ &  $-6.3\pm3.2$ &  $0.0\pm 3.5$ \\
12&$4.8\pm8.0$ &  $0.0\pm 7.0$ &  $0.0\pm7.6$ &  $1.5\pm 7.7$ \\
13&$3.0\pm4.3$ &  $-5.9\pm 4.3$ &  $-2.1\pm4.4$ &  $14.1\pm 4.3$ \\
14&$-4.5\pm4.4$ &  $-2.3\pm 4.5$ &  $4.1\pm4.3$ &  $3.2\pm 4.3$ \\
15&$0.0\pm3.6$ &  $4.6\pm 3.4$ &  $0.6\pm3.2$ &  $4.9\pm 3.3$ \\
\hline\hline
\end{tabular}
\end{center}
\end{table*}
\vfill \eject

\begin{table*}
\caption{The experimental values of the $S-$coefficients for the 15
observation periods of Ref.\cite{peters}.}
\begin{center}
\begin{tabular}{cllll}
\hline\hline Observation~$i$& $S_{s1}[{\rm x}10^{-16}]$ &
$S_{c1}[{\rm x}10^{-16}]$ & $S_{s2}[{\rm x}10^{-16}]$ &
$S_{c2}[{\rm x}10^{-16}]$   \\
\hline
1&$11.2\pm4.7$ &  $11.9\pm 4.9$ &  $1.8\pm4.9$ &  $0.8\pm 4.5$ \\
2&$1.8\pm6.5$ &  $-4.3\pm 6.5$ &  $6.4\pm6.4$ &  $1.8\pm 6.4$ \\
3&$-3.3\pm3.8$ &  $2.9\pm 3.8$ &  $-5.9\pm3.8$ &  $4.6\pm 4.0$ \\
4&$12.7\pm5.1$ &  $14.3\pm 5.5$ &  $-1.9\pm5.3$ &  $-3.3\pm 5.1$ \\
5&$4.7\pm8.4$ &  $-6.9\pm 7.3$ &  $-1.8\pm8.0$ &  $-7.8\pm 7.0$ \\
6&$5.2\pm5.8$ &  $-3.0\pm 5.9$ &  $7.1\pm5.9$ &  $-5.9\pm 5.8$ \\
7&$11.1\pm5.3$ &  $-13.4\pm 5.4$ &  $-4.5\pm5.5$ &  $-9.8\pm 5.5$ \\
8&$-12.1\pm8.9$ &  $0.0\pm 8.8$ &  $-3.1\pm9.0$ &  $1.4\pm 8.9$ \\
9&$-4.8\pm6.3$ &  $6.5\pm 6.4$ &  $-8.1\pm6.3$ &  $3.5\pm 6.5$ \\
10&$9.8\pm5.0$ &  $4.8\pm 5.0$ &  $1.9\pm5.0$ &  $-9.2\pm 4.8$ \\
11&$0.0\pm3.2$ &  $-3.9\pm 3.6$ &  $1.0\pm3.1$ &  $-2.2\pm 3.4$ \\
12&$-12.7\pm7.7$ &  $8.5\pm 6.8$ &  $-8.3\pm7.2$ &  $-7.1\pm 7.4$ \\
13&$-7.9\pm4.7$ &  $-4.3\pm 4.8$ &  $-1.9\pm4.8$ &  $-6.2\pm 4.7$ \\
14&$16.1\pm4.9$ &  $12.0\pm 5.2$ &  $2.9\pm4.9$ &  $-9.6\pm 4.8$ \\
15&$13.9\pm3.9$ &  $-7.0\pm 3.4$ &  $-3.3\pm3.5$ &  $3.0\pm 3.6$ \\
\hline\hline
\end{tabular}
\end{center}
\end{table*}

\section{Analysis of the data}

The analysis of Ref.\cite{noi} was restricted to the combinations
\BE \label{csid}
      {C}_{11}\equiv \sqrt{{C}^2_{s1}
      + {C}^2_{c1}}
\EE
\BE \label{c2sid}
      {C}_{22}\equiv \sqrt{{C}^2_{s2}
      + {C}^2_{c2}}
\EE
 \BE \label{ssid}
      {S}_{11}\equiv \sqrt{{S}^2_{s1}
      + {S}^2_{c1}}
\EE
\BE \label{s2sid}
      {S}_{22}\equiv \sqrt{{S}^2_{s2}
      + {S}^2_{c2}}
\EE
This was useful to reduce the model dependence in the analysis of
the data. In this way, in fact, the right ascension ${\alpha}$ and
any possible uncertainty related to the definition of the sidereal
time drop out from the theoretical predictions that will only depend
on $|{\gamma}|$, and the overall normalization
 $|K|$. The relevant numbers for these auxiliary quantities can be found in
Table 3.
\begin{table*}
\caption{ The experimental values of the combinations of $C-$ and
$S-$ coefficients defined in Eqs.(\ref{csid})-(\ref{s2sid}) as
obtained from our Table 1 and Table 2. }.
\begin{center}
\begin{tabular}{cllll}
\hline\hline Observation~$i$ &$ {C}_{11}  [{\rm x}10^{-16}]$ & $
{C}_{22} [{\rm x}10^{-16}]$ & $ {S}_{11}  [{\rm x}10^{-16}]$ &
$ {S}_{22}  [{\rm x}10^{-16}]$ \\
\hline
1&$5.9\pm4.7$ &  $3.5\pm 4.6$ &  $16.3\pm4.8$ &  $2.0\pm 4.9$ \\
2&$20.6\pm6.4$ &  $12.5\pm 6.5$ &  $4.6\pm6.5$ &  $6.6\pm 6.4$ \\
3&$5.3\pm3.6$ &  $5.3\pm 3.6$ &  $4.4\pm3.8$ &  $7.5\pm 3.8$ \\
4&$6.1\pm4.6$ &  $9.0\pm 4.8$ &  $19.1\pm5.3$ &  $3.8\pm 5.1$ \\
5&$2.4\pm8.4$ &  $16.5\pm 8.0$ &  $8.4\pm7.7$ &  $8.0\pm 7.1$ \\
6&$5.5\pm5.9$ &  $11.0\pm 5.9$ &  $6.0\pm5.9$ &  $9.2\pm 5.9$ \\
7&$9.5\pm5.7$ &  $6.5\pm 5.5$ &  $17.4\pm5.4$ &  $10.7\pm 5.5$ \\
8&$11.0\pm7.9$ &  $18.7\pm 7.9$ &  $12.1\pm8.9$ &  $3.4\pm 9.0$ \\
9&$9.1\pm6.5$ &  $13.1\pm 6.4$ &  $8.1\pm6.4$ &  $8.8\pm 6.4$ \\
10&$7.0\pm4.8$ &  $6.5\pm 4.7$ &  $10.9\pm5.0$ &  $9.4\pm 4.8$ \\
11&$6.4\pm3.1$ &  $6.3\pm 3.2$ &  $3.9\pm3.6$ &  $2.4\pm 3.4$ \\
12&$4.8\pm8.0$ &  $1.5\pm 7.7$ &  $15.3\pm7.4$ &  $10.9\pm 7.3$ \\
13&$6.6\pm4.3$ &  $14.3\pm 4.3$ &  $9.0\pm4.7$ &  $6.5\pm 4.7$ \\
14&$5.1\pm4.5$ &  $5.2\pm 4.3$ &  $20.0\pm5.0$ &  $10.0\pm 4.8$ \\
15&$4.6\pm3.4$ &  $5.0\pm 3.3$ &  $15.6\pm3.8$ &  $4.4\pm 3.5$ \\
\hline\hline
\end{tabular}
\end{center}
\end{table*}
Thus, we obtain the relations
\BE \label{sid1}
      {C}_{11}= {{1}\over{4}} |K|
      \sin 2|\gamma | \sin 2\chi
~~~~~~~~~~~~~~~~~~~
      {C}_{22}= {{1}\over{4}} |K|\cos^2\gamma
      (1+ \cos^2\chi) .
\EE
The corresponding ${S}$-coefficients are also predicted as
${S}_{11}= {C}_{11}/\cos\chi$ and ${S}_{22} = {{2\cos\chi}\over{1+\cos^2\chi}}$ ${C}_{22}$. Using the weighted
averages of the values in Table 3 for these coefficients  \BE
\langle{C}_{11} \rangle= (6.7 \pm 1.2)\cdot 10^{-16}~~~~~~
\langle{C}_{22} \rangle= (7.6 \pm 1.2)\cdot 10^{-16} \EE \BE
\langle{S}_{11} \rangle= (11.0 \pm 1.3)\cdot 10^{-16}~~~~~~
\langle{S}_{22} \rangle= (6.3 \pm 1.3)\cdot 10^{-16} \EE  one gets
\cite{noi} an average
 declination \BE \label{gammamean}|\gamma|\sim  43^o \pm 3^o \EE and an average
 normalization
\BE \label{kappa2} |K|\sim (33\pm 3)\cdot 10^{-16}\EE.

To check the regularity of the data, one can also extract the
average declination angle from various groups of observation
periods. Packing the data in groups of three observations, one gets
the results shown in Table 4. Notice the remarkable consistency
among the various sets of data.
\begin{table*}
\caption{ The absolute value of the declination angle $|\gamma|$ and
the normalization factor $|K|$, obtained from the data in Table 3
for groups of three observation periods.}
\begin{center}
\begin{tabular}{cll}
\hline\hline Observations  & $ |\gamma|$~[degrees] & $ |K| [{\rm
x}10^{-16}]$  \\
\hline
(1-3)   &  ${46}^{+7}_{-9}$  &  $29\pm 6$ \\
(4-6)   &  ${32}^{+7}_{-9} $  & $33\pm 7$ \\
(7-9)   &  ${39}^{+6}_{-8}  $  &  $43\pm 7$\\
(10-12) &  ${42}^{+7}_{-9}$ &  $ 27\pm 5$ \\
(13-15) &  ${41}^{+6}_{-7}$  &  $ 36\pm 5$ \\
\hline\hline
\end{tabular}
\end{center}
\end{table*}

%\begin{figure} [ht]
%\epsfig{figure=Figure1.eps,height= 8.8true cm,width=15.5 true
%cm,angle=0} \caption{ The absolute value $|\gamma|$ for each
%observation period of Ref.\cite{peters} as obtained from the
%quantities $C_{11},C_{22},S_{11},S_{22}$ defined in Eqs.(10)-(13).}
%\end{figure}
%\vskip 20pt
To extend the above analysis and get information on $\alpha$ and the
sign of $\gamma$, we shall now try to re-construct
 the full amplitudes $S(t)$ and $C(t)$ of Eq.(\ref{basic2}) or,
more precisely, their variable parts $S(t)-S_0$ and $C(t)-C_0$. In
this way, in fact, possible problems related to the deconvolution of
the signal in its elementary components drop out. To this end, we
shall use Eqs.(\ref{amorse1}) and (\ref{amorse2})to generate a
suitable signal where:

~~~~~~i) the individual $C-$ and $S-$ coefficients are fixed to
their experimental values of Ref.\cite{peters} reported in our
Tables 1 and 2

~~~~~ii) the average times for the 15 individual observation periods
are taken from Fig.3 of Ref.\cite{peters}. Once the figure is
reproduced on the screen, these time coordinates can be extracted to
a good accuracy from the number of pixels ${\rm Npixel}$ of the data
points. In this way we get (in days since 1/1/2000) $t= {{({\rm
Npixel}-475)}\over{1.7}} +1900$ where ${\rm Npixel} $= 317, 344,
351, 365, 377, 400, 412, 423, 427, 437, 470, 493, 496, 503, 507 for
the 15 observation periods.

The resulting data sets are reported in Table 5.\vfill\eject

\begin{table*}
\caption{The re-constructed amplitudes obtained following the steps
i) and ii) described in the text.}
\begin{center}
\begin{tabular}{cll}
\hline\hline Observation $i$ & $(C(t_i)-C_0)[{\rm
x}10^{-16}]$ & $(S(t_i)-S_0)[{\rm x}10^{-16}]$  \\
\hline
$1$ &  $2.7\pm 6.8$ &  $17.0\pm6.9$ \\
$2$ &  $19.0\pm 9.2$ &  $-7.6\pm9.2$  \\
$3$ &  $9.3\pm 5.5$ &  $0.9\pm5.5$ \\
$4$ &  $13.3\pm 6.8$ &  $12.1\pm7.5$ \\
$5$ &  $-13.6\pm 12.2$ &  $11.7\pm 11.3$ \\
$6$ &  $-7.7\pm 8.3$ &  $-13.3\pm8.3$ \\
$7$ &  $-7.6\pm 8.1$ &  $-20.7\pm7.8$ \\
$8$ &  $-2.4\pm 11.7$ &  $-3.0\pm12.6$ \\
$9$ &  $3.1\pm 9.5$ &  $3.4\pm9.1$ \\
$10$ &  $12.0\pm 6.8$ &  $-6.3\pm7.1$  \\
$11$ &  $0.6\pm 4.9$ &  $-4.5\pm4.9$ \\
$12$ &  $-1.9\pm 11.3$ &  $-11.1\pm10.5$ \\
$13$ &  $4.1\pm 6.2$ &  $-0.2\pm6.8$  \\
$14$ &  $2.9\pm 6.4$ &  $-18.8\pm7.0$  \\
$15$ &  $-1.4\pm 5.0$ &  $-14.7\pm5.1$  \\
\hline\hline
\end{tabular}
\end{center}
\end{table*}
We can now try to fit the values of Table 5 to the theoretical
predictions ($\mu=\pm 1$) \BE
 C(t)-C_0  = \mu|K|
 ( {{\sin2\chi
\sin2\gamma\cos(\alpha-\tau)}\over{4}}
+{{(1+\cos^2\chi)\cos^2\gamma\cos2(\alpha-\tau)}\over{4}})\EE
 \BE
S(t)-S_0=\mu|K|({{\sin\chi\sin2\gamma\sin(\alpha-\tau)}\over{2}}
+{{\cos\chi\cos^2\gamma\sin2(\alpha-\tau)}\over{2}})\EE to obtain
information on $|K|$, $\alpha$ and the sign of $\gamma$.

Before presenting the results, we observe that the structure of the
problem is such that one should first suitably constrain the fit. In
fact, all together there are 16 possible choices arising from the 2
possible signs of $\mu$, the 2 possible signs of $\gamma$ and the 4
possible choices of $\alpha$ ($\sin\alpha=\pm |\sin\alpha|$ and
$\cos\alpha=\pm|\cos\alpha|)$. Considering all possibilities we have
found that the various fits to the values of $C(t)-C_0$ do not
provide any definite information. In fact, in all cases the quality
of  the chi-square is very close to that of the "null result"
defined by $|K|=0$ and the parameters cannot be constrained in any
meaningful way.

Considering the values of $S(t)-S_0$ the situation is different. In
fact the fit routine (MINUIT) finds in this case two configurations
whose chi-square is definitely lower than the chi-square of the
"null-result" ($\sim 38$). The existence of multiple solutions had
to be expected since the $C-$ (and $S-$) coefficients of Eqs.(4)-(8)
are unchanged under the simultaneous replacements $\gamma \to
-\gamma$ and $\alpha\to \alpha +\pi$. The absolute minimum, whose
chi-square is $\sim 16$ (for 12 degrees of freedom), corresponds to
$\mu=-1$ with the physical parameters being in the following ranges
\BE \label{fit}
 |K|=(39\pm 15)\cdot10^{-16}~~~~~~~~~~~~\gamma=-{30^o}^{
+16^o}_{-22^o}~~~~~~~~~~~~\alpha=204^o\pm 12^o\EE As one can see,
the values for $|K|$ and $|\gamma|$ from our reconstructed amplitude
for $S(t)-S_0$ are in good agreement with Eqs.(\ref{gammamean}) and
(\ref{kappa2}), and with those reported in our Table 4, that were
deduced from the average values of the positive-definite
coefficients $C_{11}$, $C_{22}$, $S_{11}$ and $S_{22}$.

We have no explanation for the different behaviour obtained using
our re-constructions of $C(t)-C_0$ and $S(t)-S_0$. Perhaps this
might be due to the asymmetric experimental set-up where only one
cavity is rotated or, since the 15 data sets, each spanning from 24
hours to 100 hours in length, have been summarized into just a
single point, the inherent inexactness in our re-construction
affects differently the two sets of data.  On the other hand, the
indications we have obtained from $S(t)-S_0$ are so consistent with
the previous results of Ref.\cite{noi}, and with those reported in
our Table 4, to suggest that they should still persist in a more
refined analysis using the full raw data or after inclusion of new
observations.

\section{Summary and conclusions}

The possibility of a large concentration of matter in the region of
sky $\gamma\sim -44^o$ and $\alpha \sim 202^o$, the "Great
Attractor", was proposed to describe the deviations of galaxies from
a pure Hubble flow \cite{lynden88}-\cite{burstein} in our local
Universe. Its inclusion in multi-attractor models, see e.g.
Ref.\cite{marinoni}, provides a successful representation of the
observed velocity field for large samples of galaxies, such as the
$\sim 3400$ ones contained in the MARK III catalog \cite{willick}.

The resulting motion of the Local Group is known to exhibit some
discrepancy with respect to the direction obtained from a dipole fit
to the anisotropy of the CMB. The value of the misalignment angle is
both sample- and model dependent. For instance in the
multi-attractor model of Ref.\cite{marinoni} it ranges from $11^o$
to $49^o$ with a typical value of $\sim 30^o$.

In principle, the existence of this discrepancy might also show up
in ether-drift experiments. Up to now, these have been analyzed
assuming that the kinematical parameters of a hypothetical Earth's
"absolute motion" should coincide with those extracted from a dipole
fit to the COBE data. Rather, in our opinion, one should leave out
the angular variables $\gamma$ and $\alpha$ and the quantity
$K=(1/2-\beta+\delta){{v^2}\over{c^2}}$ as free parameters and check
the overall agreement with the data.

In this paper, using the extensive ether-drift observations reported
by Herrmann et al. \cite{peters}, we have obtained evidence for an
angular pair that sizeably differs from the values $\alpha\sim
168^o$ and $\gamma \sim-6^o$ obtained from a dipole fit to the
anisotropy of the CMB. The motion we have detected points toward
$\gamma=-{30^o}^{ +16^o}_{-22^o}$ and $\alpha=204^o\pm 12^o$ and the
central values coincide with those of the Shapley supercluster
\cite{shapley}. However, the error in $\gamma$ is large. Therefore,
we could use the result of Ref.\cite{noi} $|\gamma|=43^o\pm 3^o$, to
further sharpen our estimate. In this way, the angular parameters of
the Earth's motion coincide almost exactly with those of the core of
the Great Attractor ($\gamma\sim -44^o$ and $\alpha\sim 202^o$).

Actually, the agreement with the position of the Great Attractor is
even too good. In fact, an ether-drift observation on the Earth
contains, in principle, the effect of the solar motion (`sm')
relatively to the centroid of the Local Group. For instance, using
for this motion the values of Ref.\cite{vandenbergh}, $v_{\rm
sm}\sim (306 \pm 18) $ km/s, $\gamma_{\rm sm} \sim (51^o\pm 6^o) $,
$\alpha_{\rm sm}\sim (332^o\pm 11^o)$, and using for the peculiar
motion of the Local Group the values of Ref.\cite{marinoni} obtained
fitting the multi-attractor parameters to the full MARK III catalog,
namely ${V}_{\rm LG}\sim 488 \pm 80$ km/s (in the CMB frame),
${\gamma}_{\rm LG}\sim -50^o\pm 8^o$ and ${\alpha}_{\rm LG}\sim
202^o\pm 10^o$, one can predict a theoretical velocity through \BE
{\bf v}_{\rm th}={\bf V}_{\rm LG} + {\bf v}_{\rm sm} \EE obtaining
the following values \BE \label{th} v_{\rm th}= (276 \pm 71)~{\rm
km/s}~~~~~~ \gamma_{\rm th}= {-30^o}^{+18^o}_{- 23^o}~~~~~~
\alpha_{\rm th}= {240^o}^{+21^o}_{-28^o} \EE These might be compared
with the results of our fit (\ref{fit}) and with our
Eqs.(\ref{gammamean}) and (\ref{kappa2}) having a theoretical
estimate of $|(1/2-\beta+\delta)|$ to transform into a velocity
value, through Eq.(\ref{cappa}), the precise indication
(\ref{kappa2}) on the normalization factor $|K|$.

To this end, we observe preliminarily that our model-independent
analysis leads to rather large values of the RMS anisotropy
parameter. In fact, using our result $|K|\sim (33\pm 3)\cdot
10^{-16}$ and the theoretical range $v_{\rm th}\sim (276\pm 71)$
km/s reported above, one obtains a range of the RMS parameter
$25\cdot 10^{-10}\leq |(1/2-\beta+\delta)|\leq 70\cdot 10^{-10}$
(with a central value $39\cdot 10^{-10}$) in good agreement with the
theoretical prediction $|(1/2-\beta+\delta)|_{\rm th}\sim 42\cdot
10^{-10}$ of Refs.\cite{pagano,pla}. Equivalently, using the two
theoretical inputs $|(1/2-\beta+\delta)|_{\rm th}\sim 42\cdot
10^{-10}$ and $v_{\rm th}\sim (276\pm 71)$ km/s one predicts
$20\cdot 10^{-16}\leq |K|_{\rm th}\leq 56\cdot 10^{-16}$ (with a
central value $36\cdot 10^{-16}$) in good agreement with our Eq.
(\ref{kappa2}).

This should be compared with the result of Ref.\cite{peters} (where
the values $v\sim 370$ km/s, $\alpha\sim 168^o$ and $\gamma\sim
-6^o$ were assumed) $|(1/2-\beta+\delta)|\sim (2\pm 2)\cdot
10^{-10}$. In this way, one would predict $|K|\sim (3\pm 3)\cdot
10^{-16}$ which is one order of magnitude smaller than the value
reported in our Eq.(\ref{kappa2}). In this sense, our results show
that the very small RMS parameter reported in Ref.\cite{peters} ,
rather than being due to the smallness of the signal, might
originate from more or less accidental cancellations among the
various entries.

\vskip 30 pt \centerline{\bf{Acknowledgements}} We thank C. Marinoni
for kindly providing the estimate of the errors on the velocity
parameters reported in Table 6 of Ref.\cite{marinoni}
 \vfill\eject

\end{document}